\documentclass{sig-alternate-10pt}

\usepackage{pseudocode}
\newcommand{\mn}{\mathbb{N}} 

\newcommand{\Probsym}{\mathbb{P}}
\newcommand{\Prob}[1]{\Probsym\left[#1\right]}

\begin{document}

\title{Towards Informative Statistical Flow Inversion}


\numberofauthors{1} 
%
\author{
%
%
Richard Clegg, Hamed Haddadi, Raul Landa, Miguel Rio \\ 
       \affaddr{University College London}\\ 
       \affaddr{United Kingdom}\\
       \email{[rclegg|hamed|rlanda|mrio]@ee.ucl.ac.uk}\\
}

\maketitle
\begin{abstract}
A problem which has recently attracted research attention is that
of estimating the distribution of flow sizes in internet traffic.
On high traffic links it is sometimes impossible to record every
packet.  Researchers have approached the problem of estimating
flow lengths from sampled packet data in two separate ways. 
Firstly, different sampling methodologies can be tried to more
accurately measure the desired system parameters.  One such
method is the {\em sample-and-hold\/} method where, if a packet
is sampled, all subsequent packets in that flow are sampled.
Secondly, statistical methods can be used to ``invert'' the sampled
data and produce an estimate of flow lengths from a sample.

In this paper we propose, implement 
and test two variants on the sample-and-hold method.  In addition we 
show how the sample-and-hold method can be inverted to get an 
estimation
of the genuine distribution of flow sizes.  
Experiments are carried out on real network traces to compare standard
packet sampling with three variants of sample-and-hold.  The methods
are compared for their ability to reconstruct the genuine distribution
of flow sizes in the traffic.

\end{abstract}




\section{Introduction}

Routers at the core of the internet deal with millions of packets per 
second on multiple interfaces. From a network 
operations perspective, it is vital for the administrators to be aware 
of the volume and types of the packets that are traversing their 
networks. In order to achieve this objective, routers are required to 
collect management information but it is impossible to keep a record 
of all the packets. Thus, given the vast amount of information that needs to be 
collected, routers {\em sample\/} the traffic stream. This means
that only a subset of the packets traversing any interface of the router are
processed. Today, the most commonly implemented technique is {\em
packet 
sampling\/}, where 1 packet out of every $N$ is chosen on a random or periodic
basis, and integrated into a flow record in the router memory.

In many practical cases, $1/N$ packet sampling is followed by
multiplication of the recovered statistics by $N$ ({\em N-multiplication}). 
This simple technique can
be used to recover a number of {\em packet level statistics\/} of interest.  For
example, the number of SYN packets, TCP packets, ICMP packets or packets
to or from given destinations in the original trace can be estimated by
this process.  However, the distribution of flow lengths can not be recovered
by this procedure (see Section \ref{sec:packetsamp}).

The problem at the heart of this paper is that of recovering the distribution
of flow lengths from sampled data.
The {\em flow inversion\/} problem amounts to mathematically compensating 
for the effects of 
sampling in order to estimate the distribution of flow lengths which would have
been observed in the original data. 
There has been great
research activity around the flow distribution inversion 
\cite{fisher,duffield,inverting} and this is discussed in
Section \ref{sec:other_work}. Based on 
the previous studies on analysis of the NetFlow performance 
\cite{choi}, it is evident that {\em N-multiplication} upsets the
flow level statistics of the original data stream \cite{haddadi}.

\subsection{Outline}

This paper focuses on considering sampling methods and their 
use to estimate the flow length distribution in real traffic.
Packet sampling has many useful statistical properties, 
but it is a hard problem to recover the flow distribution from packet-sampled
data. This is sometimes called the {\em flow inversion problem}.
In this paper we investigate three techniques based on the idea of
the sample-and-hold method \cite{estan}.  This method
was originally conceived to track the largest flows
in the traffic (with applications related to billing) \cite{billing}.

Section \ref{sec:packetsamp} gives some basic information about packet
sampling as applied in real situations.  Section \ref{sec:other_work} 
discusses other work on the problem of inferring flow distributions from
sampled data.
In Section \ref{methodology} we describe the sampling methods we use
and the inversion procedure used to recover the original flow
distribution. 
There is also a brief overview on the router memory and resources requirements.
In Section \ref{results} we have applied our proposed algorithms on packet traces 
from a backbone network and have looked at the performance of our 
algorithm.  In Section \ref{conclusion} we 
have summarized our results and discuss potential avenues for future work.

\subsection{Definitions}
\label{definitions}

The traffic on the Internet is carried in form of Internet 
Protocol (IP) packets and transmitted to the destination on a 
hop-by-hop basis by Internet routers. In order to keep account of 
the packets belonging to the same application, the concept of a flow 
is defined by router manufacturers. A {\em flow} is usually defined as a 
group of packets that have the same 5-tuple (IP protocol, source 
address, source port, destination address, destination port).
\footnote[1]{The industry sometimes uses other 
definitions such as 7-tuple format, however we choose the 5-tuple 
format which is more commonly used in the research context.}

Usually, core Internet routers carry a large number 
of flows at any given time. This pressure on the router is controlled by using 
strict rules to remove from router memory (\emph{export}) the statistics, and
thus keep the router memory buffer and CPU resources available to deal with
changes in traffic patterns by avoiding the handling of excessively large tables
of flow 
records. Cisco NetFlow \cite{netflow}, the dominant standard on 
today's routers, uses the following criteria for expiring flows in 
the cache entries: 
\begin{enumerate}
\item Flows which have been idle for a specified time 
are expired and removed from the cache (15 seconds is default).
\item Long lived flows are expired and removed from the cache (30 
minutes is default). 
\item As the cache becomes full a number of 
heuristics are applied to aggressively age groups of flows 
simultaneously. 
\item TCP connections which have reached the end of 
byte stream (FIN) or which have been reset (RST) will be expired.
\end{enumerate}
As will be seen in Section \ref{sec:terminate}, the selection of these
parameters can greatly affect the nature of the sampled traffic.

After the flow records are terminated, they are grouped together and
exported to an external aggregation point through a UDP (User datagram
Protocol) stream. The collection of these NetFlow records enables 
system administrators to have a view of general trends in spatial traffic
distribution, network host behavior, traffic matrix estimation, anomaly
detection \cite{Ringberg-Sigmetrics-07} and other relevant measurements.
However, the effectiveness of these applications is contingent upon the quality
of the flow level statics recovered from the actual network
measurements \cite{daniela}.

The {\em flow distribution\/} is the distribution of flow lengths in a 
given traffic trace.  The lengths are usually expressed in packets
but sometimes in bytes.  This can be thought of as the probability that
a given flow has a particular length.  That is, the distribution is
$(\theta_1, \dots, \theta_M)$ where
$$
\theta_i = \Prob{\text{Randomly chosen flow is of length $i$}}.
$$

\subsection{Packet sampling}
\label{sec:packetsamp}

In an analysis by Cisco \cite{netflowperf} one NetFlow-enabled access 
router used up to 68\% of its CPU on processing flow records
when an average of 65,000 flows was kept in memory. 
When sampling was used, this utilization was decreased by more than 
82\%.
There are three constraints on a core router which lead to the use 
packet sampling: 
the size of the record buffer, the CPU speed and the flow record 
look-up time. In packet sampling, in order to relax the pressure on 
the router while collecting measurements, $1$ in $N$ packets are 
chosen, and the rest are discarded.  Sometimes this is done in
a periodic way with every $N$th packet sampled.  
However, in the literature, independent and identically distributed
(iid) sampling with a fixed probability $p$ is often
considered.  The differences between periodic sampling and iid 
sampling can be important. 
Roughan \cite{roughan-comparison} has shown iid sampling is 
useful in active probing and the concepts are 
also applicable to the case of passive measurement. 

There are many advantages to iid packet sampling and it preserves 
many important characteristics of the traffic.  
However, this sampling does not
preserve the flow length distribution.  The reason for this should be 
clear but an example is illustrative.  
Imagine a situation where the flow distribution is such that
half the flows in the original trace are of length two and half are
of length one ($\theta_1 = 0.5$ and $\theta_2 = 0.5$). 
Imagine these packets are sampled in an iid manner
with $p = 0.5$.  Half of the flows of length one will be sampled but
only one quarter of the flows of length two will have both packets
sampled.  Another half of the flows of length two will have just one
packet sampled and a final quarter will have no packets sampled.
In the final sample the flow distribution will be
($\theta'_1 = 0.8$ and $\theta'_2 = 0.2$).
The problem of flow inversion is, therefore, defined as the problem
of recovering the original distribution ($\theta_i$) from the sampled traffic.

The choice of sampling strategy will have a large impact on the 
quality of the data obtained from the network. This is why, to an extent,
thereason why the 
usability of NetFlow sampled data has been questioned by researchers 
\cite{loss}. The problems with packet sampling are twofoldstem  from the
followinf effects it has on sampled flows:

\begin{enumerate}
\item {It is easy to miss short flows altogether. 
This is due to the fact that many flows be only 
a few packets long, and they may be temporally correlated. Thus, these
constituent packets may cluster together and totally evade the sampling process.
}
\item {It can be difficult to estimate the length of 
long flows. The major problem is that, for each flow, only a small subset of
packets are seen with a given probability $p$ equal to the sampling rate.
Thus, it is not clear how many packets actually were present, out of which a
given number $X_i$ were been seen.
}
\item {Flows may be mis-ranked. This means that, even though flow $A$ may seem
to be larger than flow $B$ in the sampled statistics, this is not necessarily
the case in reality \cite{ranking}. 
}
\item {Large flows may be split into smaller ones (creating sparse
flows) \cite{duffield}. This is due to the fact that some long flows have a
bursty nature, and thus may include long periods of inactivity. During these
periods, they might be mistakenly expired, and any new packets 
belonging to the same flow are mistakenly classified as part of a new flow.
}
\end{enumerate}

For applications such as billing and monitoring, the na\"{\i}ve inversion 
method of division of the final statistics by the sampling rate, or 
basically multiplying the final data by ($1/p$) will simply lead to 
inaccurate results as pointed in \cite{haddadi}. 

On the other hand, the most important problem of the current NetFlow implement
ation of packet sampling is the fact that many flows are not sampled at all, as
none of their packets are selected for sampling. In many cases, 
particularly in short-lived flows (like web and email applications, where 
a group of packets are sent together to reply a query), only one 
packet of the flow is captured. This results in NetFlow reports being 
dominated by single packet flows. 

\subsection{Practical implementation of sampling
techniques}\label{subsection:practImpl}

Even though sampling sampling techniques are used in order to simplify the
processing of data collected at a router, in practice it is a complicated
process in itself.
In a simple implementation of packet sampling in a router, there are various 
points to be considered. A core or large access Internet router must constantly
accomodate memory and processor resource constraints.
Even though it is desirable to 
keep a large number of records in the flow cache, the fast growth of this number
makes flow look-up and update a challenge. Thus, when sampling is implemented,
the operator has to decide on a few parameters.

\begin{enumerate}
\item {Sampling rate: The sampling rate has a direct impact on the 
quantity and the quality of the information formed from the data.
}
\item {Flow time-out: The length of the time-out can have an impact on 
intermittent traffic flows, such as peer-to-peer file sharing or 
Instant Messaging, where the flows may not by transmitting packets 
at full rate the whole time.
}
\item {Flow expiry: If many large flows are active, the flow cache of the
router becomes progressively full, leaving no space for new flows. 
In order to avoid this, a value must be chosen for the expiration (timeout) of
the flows.
}
\item {Flow export frequency: If this is done too frequently, 
it increases the processing load on the router. However, if it is not done
often enough, the loss of the UDP export packets in the path can effect the
quality of the gathered statistics.
}
\item {Flow cache size: The number of flows which are kept at the router 
plays a critical role on its performance. If this number is 
too large, flow look-up time becomes a difficult. On the other hand, if it 
is too small, many flows are bound to be dropped and frequent expiry and
time-out of flows will be needed.
}
\end{enumerate}

It can be observed that optimally setting all these values can be a challenging 
task for an operator. Changes in any of the above parameters can effect the
length and number of the flows which are reported by a router. In Section 
\ref{sec:terminate} we look at some of the effects of the mentioned parameters.

\subsection{Related work}
\label{sec:other_work}

Hohn and Veitch \cite{inverting} considered in some depth
the problem of producing
an estimate of flow distribution from sampled packets.  
They first looked at methods for ``inversion'' to recreate
the original flow distribution from the sampled packet data. 
They use two schemes to recreate the flow distribution from the
packet sampled data, the first based upon a binomial sum and the
second upon a Cauchy integral. These schemes can successfully
recover the flow distribution for short length flows if the sampling rate
$p$ is relatively high (for example, more than half the packets sampled
is ideal).  This is not a flaw in the methods described, but a 
fundamental limitation in the amount of information which can be retrieved
from packets sampled in this manner.

Following this, their paper proposes
a flow sampling model  which can be used in an offline analysis of flow 
records formed from an unsampled packet stream.   In this method, all
the packets are recorded and formed into flows.  Then, a subset of
these recreated flows are sampled using iid sampling with
a given probability $p$.  This sampling method proves extremely
successful at recreating the flow distribution even when the sampling
ratio $p$ is relatively small (say $p = 0.001$).
However, the intensive computing and memory requirements makes the 
implementation of such a scheme on high speed routers a challenge. 

Duffield et al \cite{duffield} have looked at recovering the flow 
length distributions from a sampled packet trace. A scaling based, 
Maximum Likelihood Estimation method is proposed and, due to its 
complexity, an iterative Expectation Maximization algorithm is tested 
on the available trace files. The biggest issue encountered by the 
authors is the complexity of the process and the adjustment of the lowest 
order weights to reflect the underlying distributions. It is 
re-established by the authors that the estimation of flow level statistics from
packet sampled data remains an open question.

In a subsequent paper, \cite{duffield-learn} the authors 
introduce threshold sampling as 
a sampling scheme that optimally controls the expected volume of 
samples and the variance of estimators over any classification of 
flows. The proposed scheme has packet capturing performed at routers, followed 
by flow formation and export and staging at a mediation station,  and 
aggregation of records at a measurement collector. 

Ribeiro et al \cite{fisher} use several methods to estimate the flow 
distribution from sampled packets.  They make use of several features 
of the TCP protocol, including the SYN flag, and the fact that 
sequence numbers can give information about the number of bytes 
between sampled 
packets.  Their work uses maximum likelihood estimators to fit a the 
distribution of flow lengths up to some maximum flow length (maximum 
flow lengths of twenty and one hundred are used in the paper).  The 
sequence numbers in particular prove helpful in extracting 
information about these short and mid-length flows.  In addition
they use a technique based on the Cram{\'e}r--Rao bound to investigate
the best possible (lowest variance)
performance of unbiased flow distribution estimators
given assumptions about the information available.  

Estan and Varghese \cite{estan} propose two algorithms for 
identifying the large flows: \emph{sample and hold} and multistage filters, 
which take a constant number of memory references per packet and use 
a small amount of memory. If $M$ is the available memory, the errors of 
the algorithms are proportional to $1/M$; by contrast, the error of an 
algorithm based on classical sampling is proportional to 
$1/\sqrt(M)$, thus providing much less accuracy for the same amount 
of memory. This scheme is intended for billing schemes where large 
flows are of higher interest to the operator.

Estan et al \cite{betternetflow} have proposed an improvement to 
NetFlow by adapting the sampling rate, enabling the router to keep a 
pre-determined number of flows in the cache. As a result of change in 
sampling rate, at each stage a normalization step is performed which 
ignores the packets that would not have been sampled if the lower sampling  
rates had been chosen. This scheme produces more concise but less accurate 
reports, due to reduction in the collected information. The constant 
change in sampling rate and renormalization stage can be an exploitable threat
to the router, allowing the degradation of its performance performance under
some attack scenarios.

Barakat et al \cite{ranking} study the possibility of detection and 
ranking of the largest flows on a link. A comparison is made between 
the blind ranking method and study how to detect and rank the largest 
flows on a link. The results indicate that at sampling rates of 
higher than 1 in a 100 it is difficult to identify the top flow with 
both methods.

Although not strictly relevant to the inversion problem it is worth
noting that there is considerable research interest in the
distribution of flow lengths in internet traffic.  This is because
the flow lengths are generally held to be heavy-tailled 
\cite{willinger1997}, that is they follow a distribution such that
\begin{equation*}
\Prob{\text{Length of flow} > x} \sim x^{- \alpha},
\end{equation*}
where $\alpha \in (0,2)$ and $\sim$ means
asymptotically proportional to as $x \rightarrow \infty$.  This means that
it is not sufficient simply to look at the flows under a given length, 
extremely long flows will also play an important part in the make-up of
the traffic.

\section{Methodology}
\label{methodology}

Our results are based on a 30 minute long trace from an 
OC-48 link on the CAIDA \cite{caida} network on $24$th April 2003. 
The trace contains 47,047,240 packets from which an average 83\% are 
TCP, 7\% are UDP. The rest are usually other network layer protocols, such
as ICMP.

The sampling strategies used in this paper are referred to as 
\begin{enumerate}
\item {\em packet sampling},
\item {\em sample-and-hold (by byte)}, 
\item {\em sample-and-hold (by packet)} and
\item {\em sample-and-hold (by SYN)}.
\end{enumerate}

Sample-and-hold (by byte) is the original sample-and-hold technique
developed in \cite{estan}.
Packet sampling as has been previously described,
is the commonly used technique of sampling each packet
in an independent manner with a given probability $p$.  This can be contrasted
with techniques which are also commonly used whereby for a given $n$, every
$n$th packet is sampled.  The three sample-and-hold techniques are
described in the next section.

\subsection{Sample-and-hold techniques}

Sample-and-hold (by byte) is a sampling technique developed in
Estan-Varghese \cite{estan}.
In this technique the router keeps track of certain flows and
samples every packet on these flows until their expiry.  The
technique was
developed with the aim of producing a sampling method in which flows that carry
greater traffic volume 
(sometimes called ``elephant'' flows)
are more likely to be ampled than smaller flows. Once a flow is expired,
due to one of the reasons 
discussed in Section \ref{definitions}, 
it is marked for export and kept 
in the router cache, until a relatively large set of flows is ready for 
export to an external aggregation point.
The process proceeds, in a packet by packet basis, as follows. When a packet is
seen which is part of a flow being tracked, that
packet is sampled.  If the packet is part of a flow which is
not being tracked, then there is a probability 
that this packet will be sampled and 
the flow will be added to the list of flows being tracked.  
Let $b$ be the length of the packet being considered in bytes.  Let $p$ be 
a constant in $(0,1)$.  The probability of starting to sample this flow
at the packet under consideration is $p_p = 1 - (1-p)^b$.  This is equivalent
to considering sampling every byte with probability $p$.

Sample-and-hold (by packet) is an obvious variant of this technique where
the probability of beginning to sample a flow at a given packet is a 
constant $p$.  This is equivalent to the technique from Estan--Varghese
but with the probability fixed rather than depending on the length in bytes of
the packet.

Sample-and-hold (by SYN) is another sample-and-hold variant based on
the Transmission Control Protocol (TCP).  A valid TCP flow is expected
to begin with exactly one packet with the SYN flag set. If a packet is not part
of the set of flows being tracked and it has 
the SYN flag set then, with probability $p$,
that packet is sampled and that flow added
to the list of flows being tracked.

The idea is that this SYN based sampling
is as close as possible to a version of
the flow-based sampling suggested by Hohn--Veitch
\cite{inverting} which can be implemented without recording every packet
and producing flows from them before sampling.
Of course, in any given traffic trace, some TCP flows
will have their SYN flag before the trace collection started.  Other
flows may have more than one SYN flag. This was observed previously by Duffield
et al 
\cite{duffield}. In their packet traces, they determined 
the proportion of those TCP flows 
containing at least one SYN packet that contained exactly one SYN 
packet.  For one data set it was 98.8\%; for another one it was 94.6\%.

In the CAIDA data investigated here, 7\% of flows which contained one
SYN packet contained at least one other.

It should be noted that not all traffic in the traces analyzed is TCP
traffic, and the Sample-and-hold (by SYN) method can 
only produce an estimate of the distribution of TCP
flows. However, we have examined our algorithm on TCP since more 
than 90\% of the traffic in our trace is TCP.

\subsection{Flow termination dilemma} \label{sec:terminate}

\begin{figure}[htbp]
	\centering
	\includegraphics[width=3.2in]{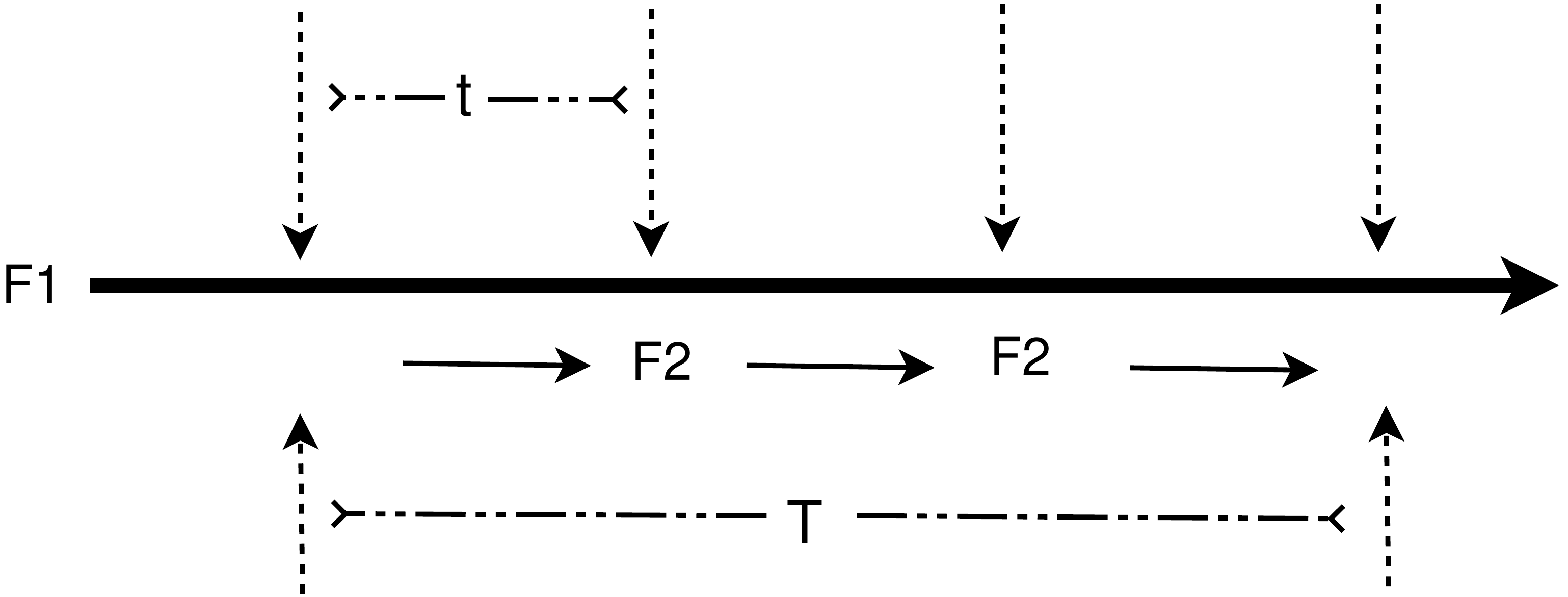}
	\caption{Changes in the flow buffer can slice up or join flows.}
\label{timeout}
\end{figure}

\begin{figure}[htbp]
	\centering
	\includegraphics[width=3.5in]{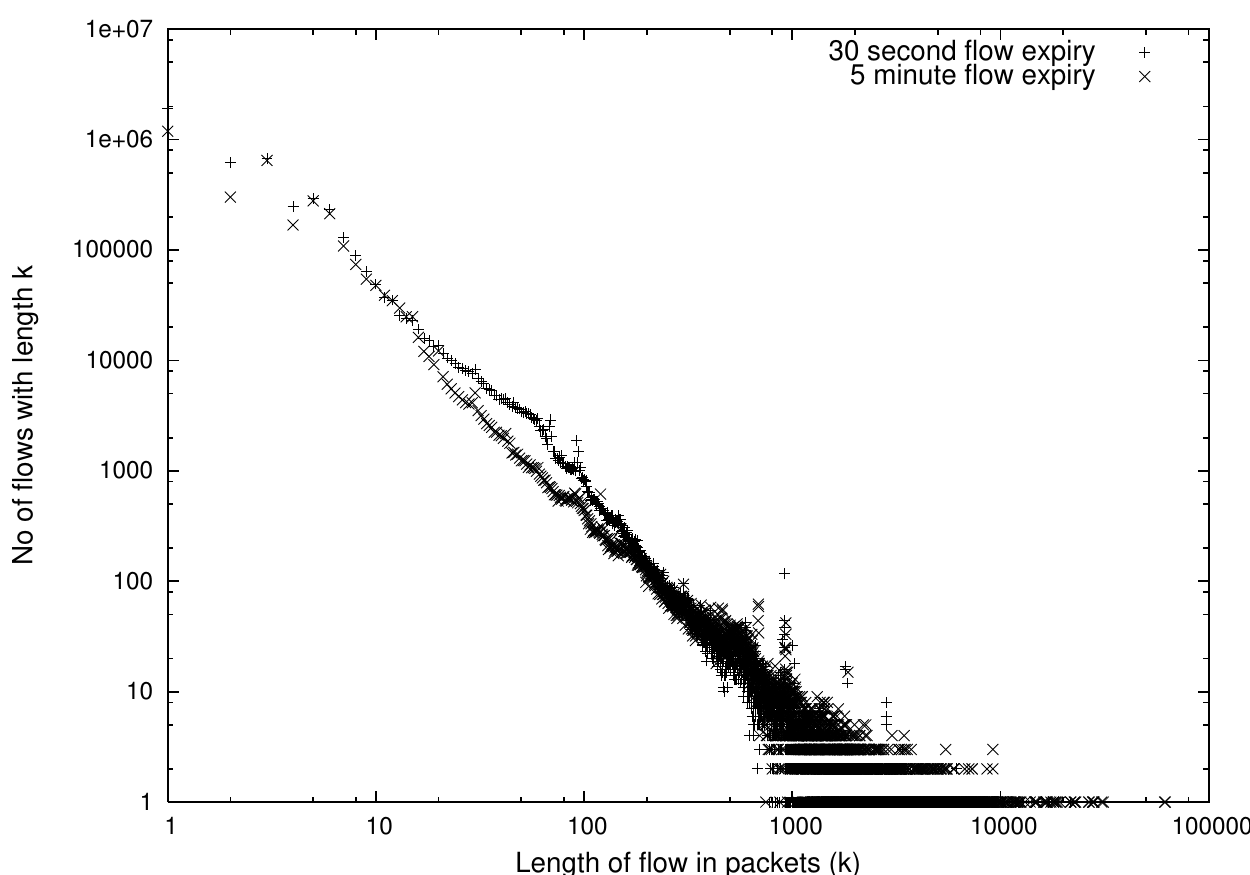}
	\caption{Number of observed flows of a given length for two
	 different timeout values on the CAIDA data.}
\label{timeout_dist}
\end{figure}

Memory constraints prevent routers from keeping flows active for long spans
of time. The flow lifetime in the router cache is configurable by the 
user. If sampling is not used, it is impossible to keep the flows in 
the buffer for more than a few minutes on a heavily utilized router. 
For example, authors in \cite{daniela} use a flow expiry timeout of 2 seconds,
which they find to be the maximum before flow loss rates reach unacceptable
levels. Figures \ref{timeout} and \ref{timeout_dist} show the effects 
of the buffer size on the accuracy of the flow reports.  It can be
seen that the longer expiry times consistently help pick out more long flows. 
Additionally, it can be seen that the distribution obtained using 
a longer expiry time is more consistent with the straight line graph
expected of a heavy-tailed flow distribution, as discussed in
Section \ref{sec:other_work}.

If the flow buffer memory in the router is chosen to be that of 
length $t$, then each section of flow $F2$ will be reported as a 
separate flow, and flow $F1$ will also be sliced up into smaller 
flows, creating so-called {\em sparse flows\/}. However if the router can 
afford to have a larger time out for the flows ($T$ in Figure 
\ref{timeout}), even if the smaller flows of $F2$ are in reality 
individual, unrelated flows (though very unlikely due to the vast 
number of source ports available for TCP packets at least), they are 
reported as a single cumulative flow. Flow $F1$ will be correctly 
reported as a complete flow.

\begin{figure}[htbp]
	\centering
\includegraphics[width=3.2in]{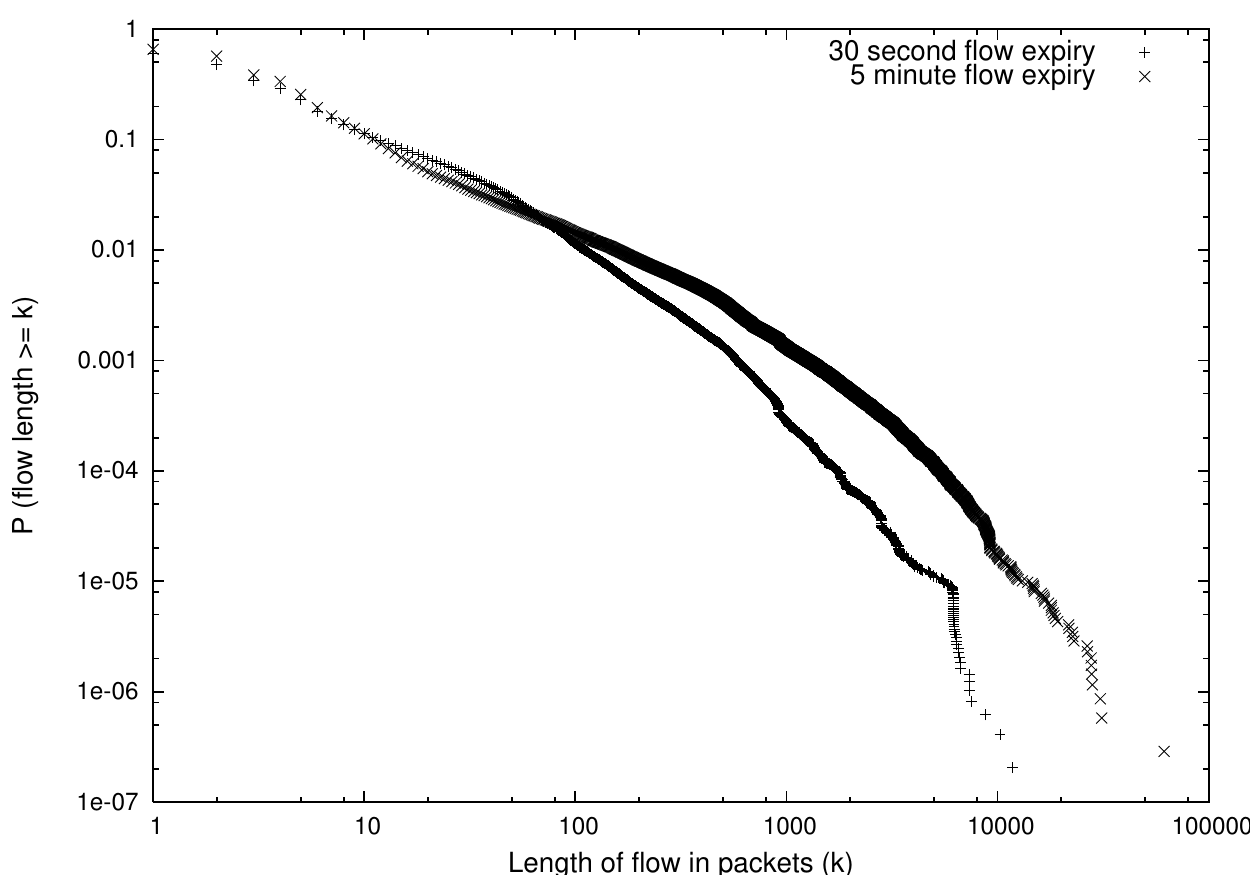}
	\caption{The complementary cumulative distribution function for
flow lengths on the CAIDA data using two different flow expiry times.}
\label{timeoutcdf}
\end{figure}

Figure \ref{timeoutcdf} displays the complimentary cumulative distribution 
function  (CCDF) of flow lengths on the CAIDA data for two different 
flow expiry lengths.  This shows clearly that, as would be expected, a 
shorter expiry time reduces the probability that the longer flows can be seen.

\subsection{Flow construction from trace files}

To build flow records out of trace files, we emulated the operation of NetFlow
on a general purpose computer, relaxing the real-time memory requirements
usually imposed on routers. Thus, we were able to greatly extend the amount of
time that detailed flow records are kept in memory, and thus construct a
baseline of unsampled measurements with which the results of our inversion
procedures could be tested. However, the sheer amount of packets in some 
high-speed Internet core traces means that we cannot process all of it in one
go. To address this, we divide time into \emph{analysis windows}, over which
flows are considered independently. For the 30 minute CAIDA datasets, however,
we only used one analysis window per trace.

The algorithm we used for the sample-and-hold
techniques of Section \ref{subsection:practImpl} is detailed in Algorithm
\ref{pseudocode:buildFlows} with
the variables explained in Table \ref{table:buildFlows}. The algorithm
for packet sampling is similar but simpler, since it does not need to track
flows.

Basically, a trace file is explored and each of its packets considered in turn.
If the packet belongs to a flow that had been previously selected for sampling,
its information is aggregated into the current flow tables. If it is not, its
flow is sampled with a probability $p$. This probability is calculated on the
basis of the sampling technique: in the case of sample-and-hold (by
packet), the same $p$ is used for all packets, while in sample-and-hold
(by byte) the probability of sampling a packet is a function of the packet
length.
If a given packet is selected, its 5-tuple $\phi$ is tracked, so that new
packets with this same $\phi$ and within the flow expiry timeout $t_t$ are
considered part of the same flow.

There are two conditions that trigger complete flow buffer exports in
Algorithm \ref{pseudocode:buildFlows}. The first one, implemented by the
boolean function {\sc FlowBufferFull()}, represents the flow buffer reaching
its maximum capacity\footnotemark.
\footnotetext{As Algorithm \ref{pseudocode:buildFlows} was
implemented for emulation, the size of the flow buffer is not a hard parameter,
but can be modified as appropriate.} 
The other one, corresponding to {\sc FlowExportTimerExpired()}, represents the
expiry of a \emph{flow analysis window}, that is, a periodic event over which
the flow collection process is restarted. 
When a buffer export occurs (independently of its triggering
condition) then the system stops tracking all flows and writes the
flow statistics to disk. This means that, for 
every $\psi \in \Psi$, the 3-tuple ($\psi$, $T_p(\psi)$, $T_b(\psi)$) is 
written to a file and the relevant data structures in program memory are
cleared.

It is informative to consider the difference between the 5-tuple $\phi$ and
the flow identifier $\psi$ in Algorithm \ref{pseudocode:buildFlows}. If the
buffer export timeout $t_w$ is significantly longer than $t_t$, it
may be possible to encounter two (or more) different flows on the same 5-tuple
$\phi$ during the same analysis window (because the flow has been timed out
and exported then seen again).  Thus, $\psi$ appends a discriminating
string to $\phi$, so that the identity of both flows is maintained. As a result
of this procedure, after a flow expires its statistics are no longer updated,
and it will be analyzed as a separate entity from other flows on the same
$\phi$. Of course, by choosing $t_t > t_w$, individual flow expiration can be
completely bypassed, and by setting $t_w$ larger than the time spanned
by the trace being analyzed, flow analysis window exports can be
completely bypassed as well. This gives Algorithm \ref{pseudocode:buildFlows}
full flexibility to explore the influence of $t_t$, $t_w$ and $N_f$ on the
sampled flow statistics and our proposed inversion techniques.

\begin{table*} [htp]
\centering
\label{table:buildFlows}
\begin{tabular}{|c|c|}
\hline
$\phi$	&	5-tuple corresponding to packet $P$\\
$t$	&	Capture time of packet $P$\\
$N_b$	&	Number of bytes in packet $P$\\
$t_s(\phi)$&	Trace time since the last packet on 5-tuple $\phi$ was seen\\
$t_t$	&	Flow expiry timeout\\
$\psi$	&	Flow identification number \\
$\Psi$	&	Set of all $\psi$ \\
$T_p(\psi)$&	Total number of packets in flow $\psi$\\
$T_b(\psi)$&	Total number of bytes in flow $\psi$\\
$p$&	Probability of starting to follow flow $\phi$\\
$t_w$	&	Flow buffer export timeout\\
$N_f$	&	Flow buffer size in records \\
\hline
\end{tabular}
\caption{Variables for the flow construction algorithm}
\end{table*}

\begin{pseudocode}{BuildFlows}{trace}
\label{pseudocode:buildFlows}

\WHILE \CALL{packetsLeft}{trace}\\  	

\DO
\BEGIN
	P \GETS \CALL{ReadPacket}{trace} \\
	(\phi, t, N_b) \GETS \CALL{DecodePacket}{P} \\	


	\\
	\IF \CALL{FlowIsBeingTracked}{\phi} \THEN \\
	\BEGIN
		\COMMENT{Has the flow expired?}\\
		\IF (t_s(\phi) > t_t) \\ 
		
		\BEGIN
		

			\psi \GETS \CALL{GetFlowID}{\phi}\\
			\CALL{TerminateFlow}{\psi} \\	
			\psi \GETS \CALL{CreateFlow}{\phi} \\
		\END \\
		
		\\
		t_s(\phi) \GETS t \\
		T_p(\psi) \GETS T_p(\psi) + 1 \\
		T_b(\psi) \GETS T_b(\psi) + N_b \\
		\\

	\END
	\ELSE \\
	\BEGIN
		\COMMENT{Is the flow going to be sampled?}\\
		\IF \CALL{flowSelectedForSampling}{p, N_b}\\

			\THEN	
			\BEGIN		


			t_s(\phi) \GETS t \\
			\psi \GETS \CALL{CreateFlow}{\phi} \\
			\Psi \GETS \psi \\
			T_p(\psi) \GETS 1 \\
			T_b(\psi) \GETS N_b \\
			\END \\
	\END\\
	
	\\
	\IF \CALL{FlowBufferFull}{|\Psi|, N_f} \\
	\OR \CALL{FlowExportTimerExpired}{t, t_w} \THEN
		\BEGIN
			\CALL{ExportFlowBuffer}{} \\
			\CALL{ResetFlowBuffer}{} \\
		\END\\


\END\\

\end{pseudocode}

For the rest of this paper we use a relatively large timeout $t_t$ of five
minutes
for the flows.  Even though this may be longer than the value usually
applied in routers (of around fifteen to thirty seconds) it helps
avoid unnecessary flow splitting. In this paper we use $t_t = t_w = 5$ 
minutes, so that all flow information is
reset every 5 minutes. After this is done, a secon post-processing step is done
where the output of this process is integrated as a single 30 miutes long file.
This is done to reduce memory consumption while avoiding dropping flows due to
lack of buffer space.

\subsection{Inverting the sampled data}

Two methods for inverting packet sampled data are given by Hohn and Veitch
\cite{inverting}.  These techniques, while mathematically sound, are 
problematic in realistic cases.  In particular, they are numerically 
unstable when estimating longer flows or rates of sampling with small
values of $p$ (where $p$ is significantly less than $1/2$).  No results
are presented here for inverting packet sampling but for an excellent
discussion of the problem, the reader is referred to \cite{inverting}.

Inverting the sample-and-hold (by packet or by byte) is, on the other hand, a
new problem.

For each flow which is not being sampled, there is a per-packet probability $p$
that the flow will start being sampled at that point. Define $q = 1-p$.  Let
$(\theta_1, \theta_2, \dots)$
be the original flow length distribution before sampling, where
$\theta_i$ is the probability that a randomly chosen flow is exactly
$i$ packets long.

Let $\theta'_i$ be the probability that the algorithm would start
sampling a randomly chosen stream $i$ packets from the end of the stream.
This is the probability that $i$ packets are sampled.  Note that
$\theta'_0 \neq 0$.
$$
\theta'_i = \begin{cases}
\sum_{j=i}^\infty p q^{j-i} \theta_j & i > 0 \\
\sum_{j=0}^\infty q^j \theta_j & i = 0
\end{cases}
$$
Let $X_i$, $i \in \mn$ be the distribution of flow lengths which can
actually be observed.  This can be thought of as the distribution
$\theta'_i$ without the probability of zero length flows.

\begin{align*}
X_i & = \Prob{\text{Sample length = $i$} | \text{Sample length $>0$}} \\
&=  \frac{\theta'_i}{\sum_{k=1}^\infty \theta'_k} \\
& = \frac{\sum_{j=i}^\infty p q^{j-i} \theta_j}
{\sum_{k=1}^\infty \sum_{j=k}^\infty p q^{j-k} \theta_j}\\
&= \frac{\sum_{j=i}^\infty q^j \theta_j}
{q^i \sum_{j=1}^\infty q^j \theta_j \sum_{k=1}^j q^{-k}} 
\end{align*}
The sum $\sum_{k=1}^j q^{-k}$ can be evaluated giving,
\begin{align*}
X_i & = \frac{(1-q) \sum_{j=i}^\infty q^j \theta_j}
{q^i \sum_{j=1}^\infty q^j (q^{-j} -1) } \\
& = \frac{(1-q) \sum_{j=i}^\infty q^j \theta_j}
{q^i \sum_{j=1}^\infty (1 - q^j)\theta_j} \\
& = \frac{(1-q) \sum_{j=i}^\infty q^j \theta_j}
{q^i[1 - \sum_{j=1}^\infty q^j \theta_j]}.
\end{align*}
Setting $i = 1$ and rearranging gives
$$
\sum_{j=1}^\infty q^j \theta_j = \frac{q X_1}{1 - q + q X_1}.
$$
Let $C = (1-q)/(1 - \frac{q X_1}{1 - q + q X_1}) = (1 - q + qX_1)$ and therefore
$$
q^i X_i = C \sum_{j=i}^\infty q^j \theta_j.
$$
Giving the final estimate 
\begin{align*}
\theta_i & = \frac{X_i - qX_{i+1}}{C} \\
& = \frac{X_i - qX_{i+1}}{1 - q + qX_1}.
\end{align*}

This method has certain obvious weaknesses.  The factor $1 - q + qX_1$ is
simply a normalization factor, the method wholly relies on the difference
between $X_i$ and $X_{i+1}$.  It is relatively insensitive to the particular
value of $p$ when $p$ is near zero (which it would be for typical sampling rates)
since the difference between $X_i - 0.99 X_{i+1}$ and $X_i - 0.999 X_{i+1}$ is
usually not great.  However, particularly at large flows this creates problems.
In particular, if $X_{i+1} > X_i$ then the method will produce a negative
estimate for the probability.  This problem can be offset to some extent
by pooling adjacent estimates so that, instead of estimating the probability
that a flow has exactly length $i$, instead an estimate is given of the
probability that the flow has a length in some range $i, i+1, \dots, i+n$.
This is discussed in the next section.

We did not find any obvious method of inverting the original Estan
sample-and-hold (by byte). The
method used in this paper is simply to assume that the data was obtained
from sample-and-hold by packet with $p$ as the probability of sampling a packet
of mean packet length using Estan's method.  

For SYN based sampling, the assumption that each TCP flow begins with exactly
one SYN flag implies that no inversion should be needed. Unfortunately, the SYN
based sampling will only sample TCP flows and can provide no information about
the distribution of UDP flows. This is a weakness of the method.

\subsection{Logarithmic binning}

When examining the flow distribution, particularly for long flows, it is
likely to be of more interest to know how many flows have a length in a given
range, rather than the number of flows with a specific length.  Therefore,
we have used a pooling technique to average data using a logarithmic
scale.  The data relating to flow lengths is averaged over bins which
contain data on a set of flow lengths (for example, one bin from all 
lengths from 1000 to 1100).  The size of the bins are chosen so that
they have a constant width (or as nearly as possible given
they are integer valued) on a logarithmic scale.
This technique is sometimes known as {\em logarithmic binning}.

Logarithmic binning is a simple way of smoothing sample data 
which is distributed on a logscale.  Let $x_k$ be the number of 
observations of a flow with length $k$.  Let $m$ be the largest flow 
observed.  The valued $x_1, x_2, \dots x_m$ will be combined into $n 
< m$ observations $X_1, X_2, \dots, X_n$.  Now, let $i_0, i_1, 
i_2, \dots, i_n$ be some
series of integers such that $i_0 < i_1 < i_2 \dots$ with $i_0 = 1$, 
$i_n$ is larger than the largest flow length observed and
$i_{k+1}/i_k$ is approximately constant for large $k$.  Now we can
derive a series $X_1, X_2, \dots, X_n$ giving the average number of
observations in the range $[i_{k-1}, i_k)$ -- note that the integer 
$i_k$ is not in this range (but will be in the range of $X_{k+1}$.
$$
X_k = \frac{\sum_{j=i_{k-1}}^{i_k -1 }x_j} {i_{k} - i_{k-1}},
$$
for $k = 1, 2, \dots, n$.
Note that, for display purposes, it makes sense to show the 
observation $X_k$ as occurring in the range $i_{k-1} - 0.5$ to $i_k - 
0.5$. 

\begin{figure}[htbp]
	\centering
\includegraphics[width=3.2in]{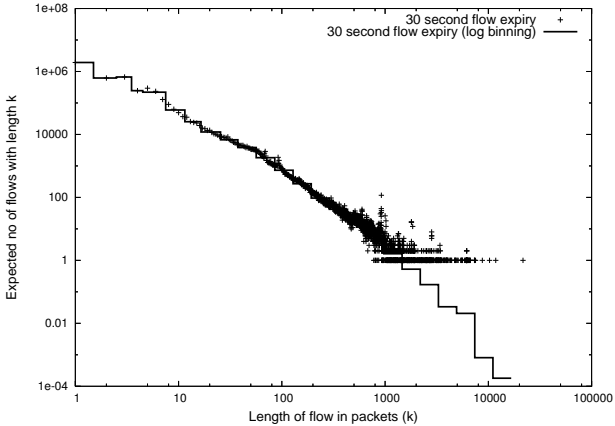}
	\caption{The effects of logarithmic binning.}
\label{fig:log_bin_demo}
\end{figure}

Figure \ref{fig:log_bin_demo} shows the results of logarithmic binning
on one of the data sets from Figure \ref{timeout_dist}.
The technique has two advantages for this study.  Firstly, it produces clearer
information for large flows.  In the large flows regime, we usually observe
either one flow or no flows, and this can make the graph on that region harder
to interpret. However, the logarithmic binning allows the graph to convey
information about how many flows are in a given range, including those
long-flow regimes where simple plots are usually uninformative. This also gives
a clearer
idea of the heavy-tailed nature of flow lengths.  Secondly, it pools those
estimators for which there is most uncertainty.  Estimating the probability that
there is a flow exactly (say) 10,005 packets long is a difficult task,
requiring vast amounts of data and processing power. On the other hand,
estimating the expected number of flows which are between 10,000 and 11,000
packets long allows the pooling of estimators to produce a more accurate
estimate using a smaller data sets, and with lesser computational demands.

\begin{figure}[htbp]
	\centering
\includegraphics[width=3.2in]{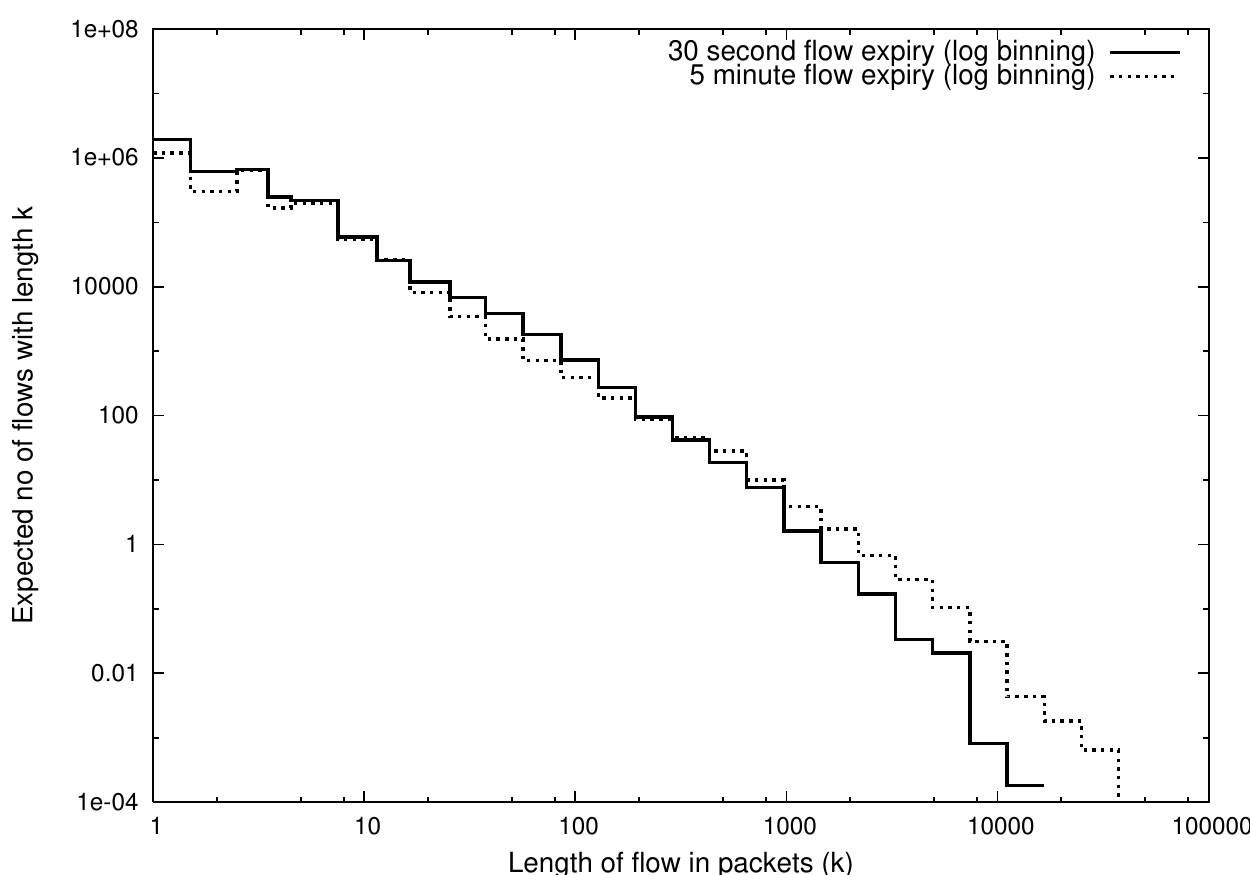}
	\caption{Figure \ref{timeout_dist} replotted with logarithmic binning.}
\label{fig:log_bin_demo2}
\end{figure}

\section{Results}
\label{results}

The results in this paper are all obtained on real network trace data.
The traces are sampled using the techniques described in the previous
section.  The flow distribution is then produced on the sampled data 
after sampling inversion techniques are applied (where such 
techniques are available) in order to recreate the original
flow distribution.  This is compared with the correct flow
distribution obtained from the unsampled data.  In order to assist
comparison, the sampling rates are chosen so that approximately 
one packet in every one hundred is sampled.  That is, the 
methods used are all sampling approximately the same percentage of
the data set and the storage requirements for each sampling method
would not be dissimilar.

The logarithmic binning method is a simple but invaluable tool for
the investigation of sampled flow length distribution.  In addition
to being a useful method for presenting the data it enables the pooling
of otherwise unreliable estimates to get a reliable estimate over a range
of values.

Packet sampling is an attractive sampling scheme for many purposes.  It 
allows recovery of many important properties of the data, however, it is
difficult to recover flow based information.  Three sample-and-hold
based schemes are used here, based upon the original sample-and-hold
described by Estan and Varghese \cite{estan} (which is here referred to
as sample-and-hold by byte).  Like packet sampling, sample-and-hold can be 
implemented in a practical setting (for example in firmware) \cite{flowmon}.

\subsection{Packet sampling}

As previously stated, inversion results are not given here for 
packet based sampling.  This is due to the extreme difficulty
of producing a flow distribution over the full range of possible
flow lengths from packet sampled data (see the discussion 
in Section \ref{sec:other_work}).  The sampling was performed to
get one packet in one hundred by setting $p= 0.01$.  This gives
$425,014$ packets sampled in $207,126$ flows, a mean of $2.1$
packets per flow.

\begin{figure}[htbp]
	\centering
	\includegraphics[width=3.2in]{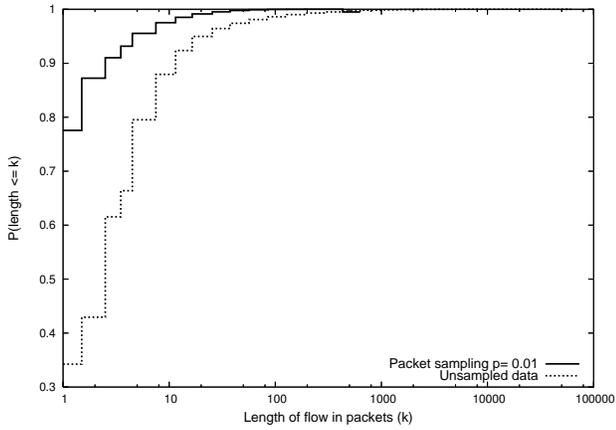}
	\caption{The impact of packet sampling with $p= 0.01$ on the flow distribution.}
\label{packetsamp_cdf}
\end{figure}

\subsection{Sample-and-hold (by packet)}

The value of $p$ was adjusted so that approximately one packet in every
one hundred was sampled.  The value of $p$ used was $0.000014$ and
this gave $413,702$ packets and $614$ flows (a mean flow length of
674 packets per flow).

\begin{figure}[htbp]
	\centering
	\includegraphics[width=3.2in]{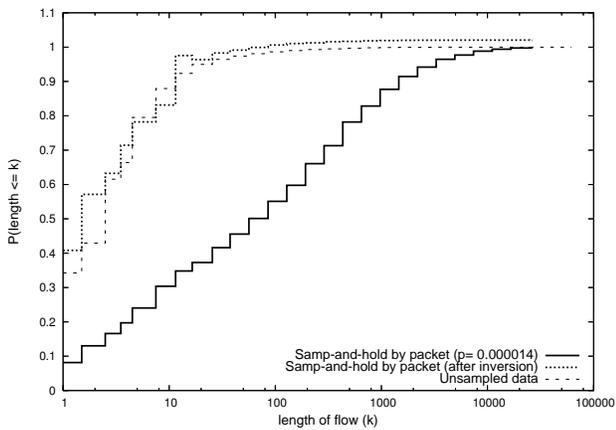}
	\caption{Cumulative density function for 
	packet based sample-and-hold with sampling of approximately one packet in 
	every one hundred on the CAIDA data.}
\label{fig:sh_1_in_100}
\end{figure}

\begin{figure}[htbp]
	\centering
	\includegraphics[width=3.2in]{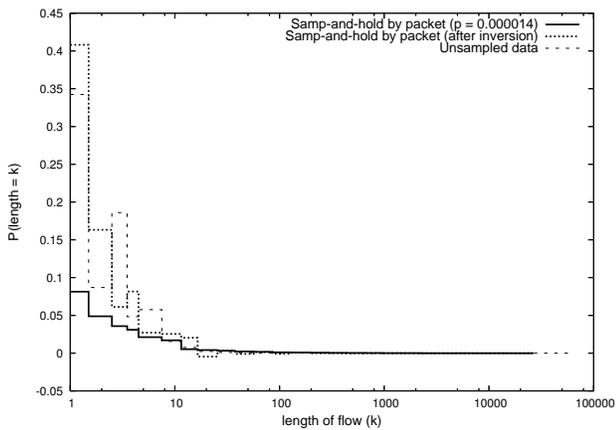}
	\caption{Density function for 
	packet based sample-and-hold with sampling of approximately one packet in 
	every one hundred on the CAIDA data.}
\label{fig:sh_1_in_100_dist}
\end{figure}

At a higher sampling rate the inversion algorithm can be seen to be
very good indeed.  With $p = 0.001$
I $10,333,134$ of $47,047,240$ packets were sampled on the CAIDA trace.
This is a very high rate of sampling but suitable for an initial test of
the sampling algorithm.

\begin{figure}[htbp]
	\centering
	\includegraphics[width=3.2in]{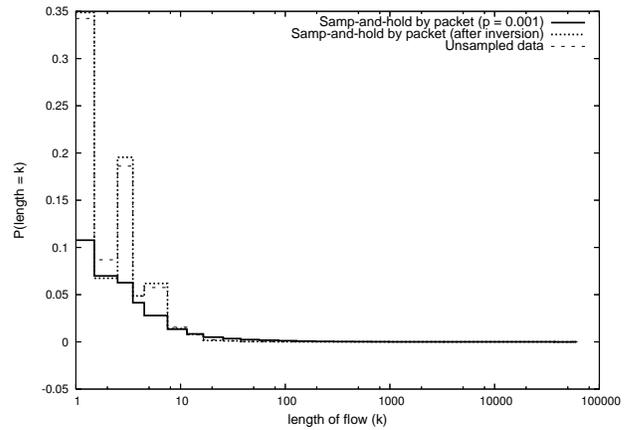}
	\caption{Density function for 
	packet based sample-and-hold with $p = 0.001$ on the CAIDA data.}
\label{fig:s_and_hold_good_dist}
\end{figure}

\begin{figure}[htbp]
	\centering
	\includegraphics[width=3.2in]{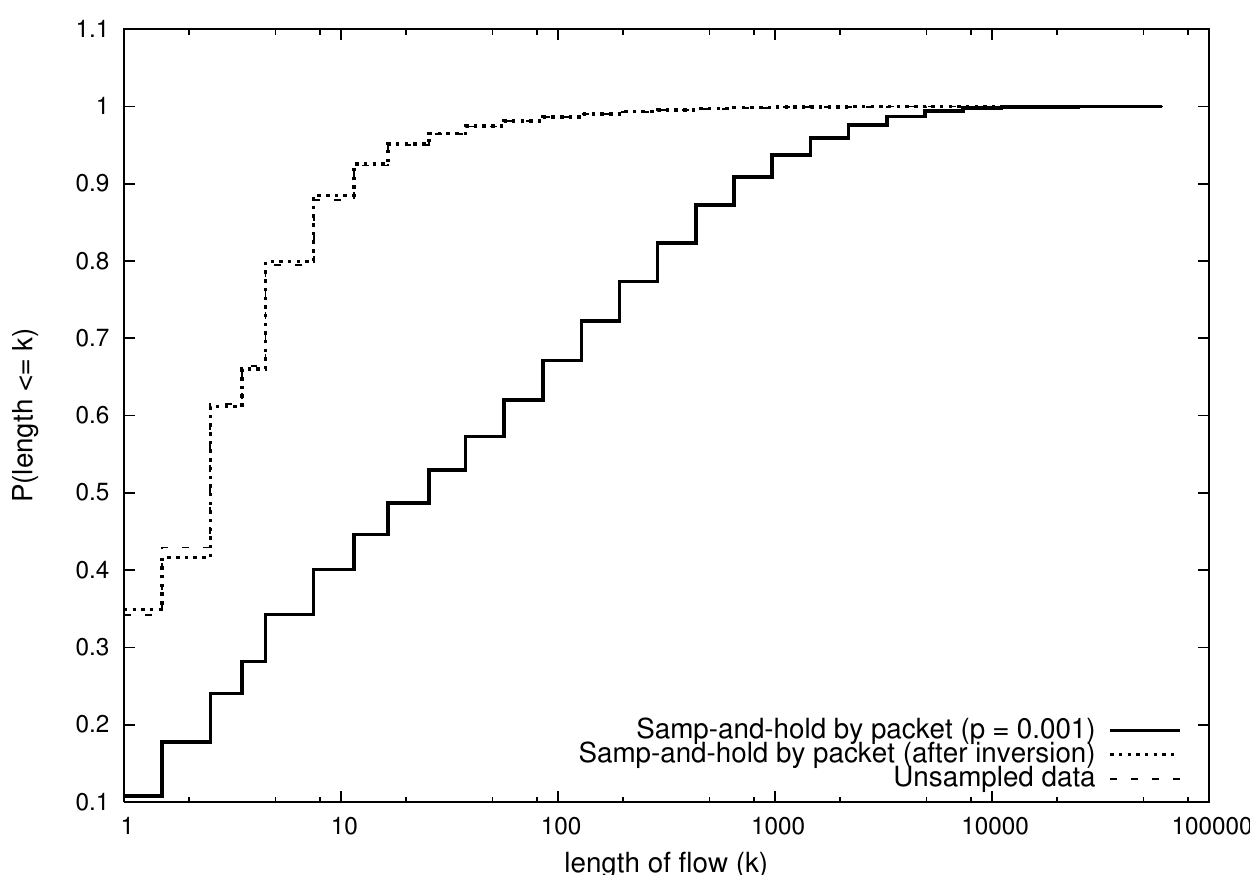}
	\caption{Cumulative density function for 
	packet based sample-and-hold with $p = 0.001$ on the CAIDA data.}
\label{fig:s_and_hold_good}
\end{figure}

Figure \ref{fig:s_and_hold_good_dist} shows the density function for this
experiment before and after inversion compared with the unsampled data.
Figure \ref{fig:s_and_hold_good} shows the distribution function for 
the same experiment.

The sample-and-hold by packet method focuses on large flows.
At the high sampling rate in Figures \ref{fig:s_and_hold_good} and
\ref{fig:s_and_hold_good_dist} the sample-and-hold by packet method inverts 
almost precisely to the original distribution.  However, nearly one 
in five packets were sampled and this is an unrealistically high sampling rate
for a highly loaded router.
At the more realistic sampling rates shown in Figures \ref{fig:sh_1_in_100}
and \ref{fig:sh_1_in_100_dist} the algorithm still performs relatively well,
particularly at higher flow lengths.  The distribution here was recovered from
only 614 sampled flows.
Another potentially useful property of the sample-and-hold by packet is that
the packets sampled can be resampled to create a sample which would have
been obtained from packet sampling with probability $p$ (where this is the
same $p$ used for sample-and-hold by packet in the first place).
This is done by sampling the first packet in each flow and 
then performing packet sampling with probability $p$ on all subsequent packets.

\subsection{Sample-and-hold (by byte)}

\begin{figure}[htbp]
	\centering
	\includegraphics[width=3.2in]{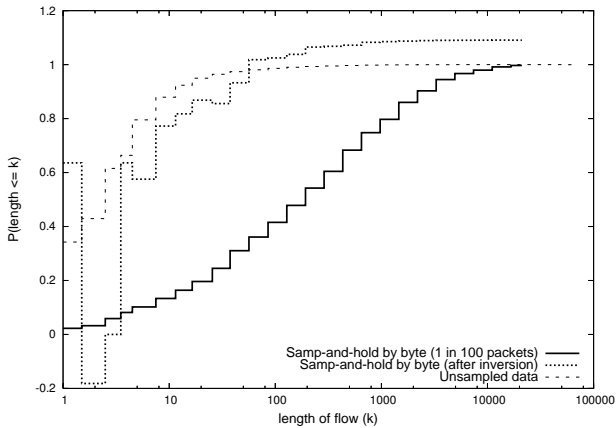}
	\caption{Cumulative density function for 
	byte based sample-and-hold with sampling of approximately one packet in 
	every one hundred on the CAIDA data.}
\label{fig:sh_1_in_100_byte}
\end{figure}

Figure \ref{fig:sh_1_in_100_byte} shows the results from sampling
the CAIDA data using the sample-and-hold (by byte) method as
proposed by Estan and Varghese \cite{estan}.  The $p$ value
has been tuned so that approximately one in one hundred packets
are sampled.  This gave $521,337$ packets sampled in total over $527$
flows, a mean flow length of $989$ packets per flow.

Sample-and-hold by byte was engineered to focus on
the largest flows (sometimes referred to in the literature as ``elephant''
flows).  It is no surprise that the flow distribution in Figure 
\ref{fig:sh_1_in_100_byte} shows this clearly.  The largest flows are tracked.
The inversion algorithm in this paper was designed to work with the packet
based sample-and-hold and it is no surprise that it does not perform
particularly well in this case.  As discussed, negative values occur in the
predicted ``probabilities" and the ``distribution" does not total to
one as it should.  This is a result of applying an algorithm which is not
quite appropriate for the data set.  Of course these issues could be fixed 
by forcing a minimum of zero and normalizing the distribution artificially.  
Nonetheless, the inverted distribution 
is an improvement over the original at longer flow lengths although performs
poorly over short flows.  The distribution in
Figure \ref{fig:sh_1_in_100_byte} is reconstructed from only 527 flows so
it is, perhaps, impressive that it is as close to the original as it is.

The advantages of sample-and-hold by byte are that it focuses clearly on
those ``elephant" flows which can dominate traffic.  It has been previously
studied in the literature and implemented in software for real sampling
applications.  On the other hand, the disadvantages are that no good 
inversion algorithm exists as yet.  In addition there is no obvious way to
recover a packet sampled data set from the sample-and-hold by byte data.

\subsection{Sample-and-hold (by SYN)}

Sample and hold by SYN was run with $p$ tuned to get approximately one
packet in every one hundred.  If the assumption of one SYN packet per
flow was correct this would simply mean setting $p=0.01$ but, as
previously discussed, this assumption is not met in the real data.
In this sample, $520,116$ packets were sampled in $68,618$ flows, a mean
of $7.6$ packets per flow.

\begin{figure}[htbp]
	\centering
	\includegraphics[width=3.2in]{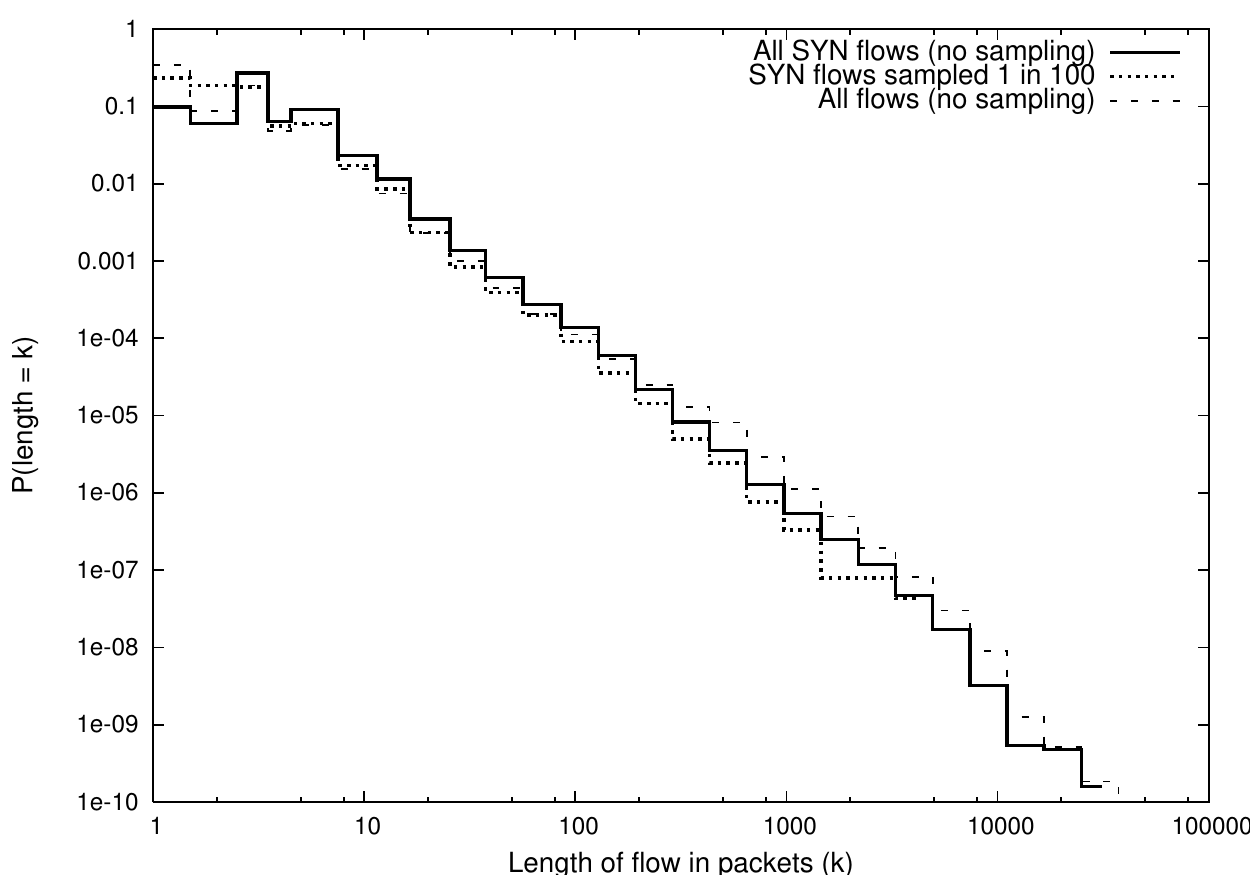}
	\caption{Cumulative density function for 
	SYN based sample-and-hold with sampling of approximately one packet in 
	every one hundred on the CAIDA data.}
\label{fig:syn_samp}
\end{figure}

\begin{figure}[htbp]
	\centering
	\includegraphics[width=3.2in]{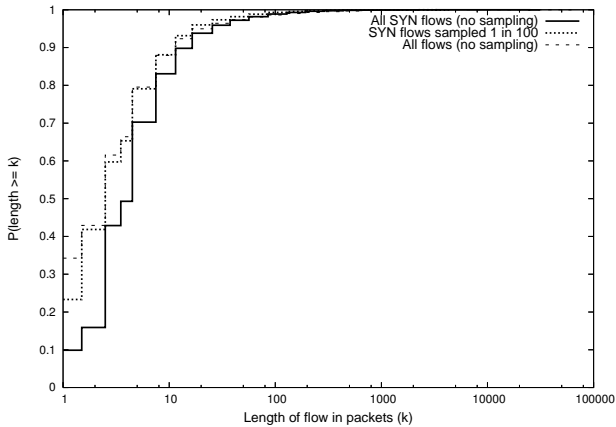}
	\caption{Cumulative density function for 
	SYN based sample-and-hold with sampling of approximately one packet in 
	every one hundred on the CAIDA data.}
\label{fig:syn_samp_cum}
\end{figure}

The sample-and-hold by SYN is actually surprisingly good at recovering
the sampled distribution as can be seen in Figures \ref{fig:syn_samp} and
\ref{fig:syn_samp_cum}.  However, this is somewhat misleading.  As can
be seen if Figure \ref{fig:syn_samp_cum}, the distribution from all SYN flows
(effectively SYN sampling at a rate of 1) is, in fact, very different from
the distribution of all the flows.  This is not unexpected.  It seems that
in this trace the non SYN flows (mostly UDP and some ICMP) are shorter and,
in particular there are more flows which are just one or two packets long.

It has already been noted that SYN sampling does not provide an unbiased
estimate of SYN flows because, in reality, flows can have more than one SYN
packet.  Many of these flows with more than one SYN packet are short flows
(perhaps because a flow with multiple SYN packets can result from trying to 
initiate a connection to a machine which is not responding).  When SYN 
sample-and-hold is used then it will be more likely to sample those flows
with multiple SYN packets (unless the sample rate is one, of course).  
The presence of such a protocol behaviour has fortuitously cancelled out the error in
the other direction and the good recovery of the flow distribution is
a product of two errors in opposite directions cancelling rather than
a true measure of the success of the algorithm.  This gives a large
element of uncertainty to the use of SYN sampling as a method for
recovering flow distributions since the basic assumption (that TCP flows
begin with a single SYN packet) is not met in real data.

A major disadvantage of sample-and-hold by SYN packet is that it
cannot provide information about non TCP packets.
A major advantage is that it gives an approximate
flow distribution with no need for inversion because it is
an approximation to flow sampling.

\section{Conclusions and future work}
\label{conclusion}

Producing flow distributions for sampled packet traces is a difficult problem.
Several authors have approached the problem of producing flow distributions
from traces sampled using standard packet sampling.  However,
different sampling methods can be used to provide a sample which makes the
recovery of the flow distribution easier while at the same time not putting
an undue requirement on memory and storage for the hardware performing the
sampling.

Of the sample-and-hold based methods studied here all have advantages and 
disadvantages.  The byte based sample-and-hold originally proposed in
\cite{estan} was intended to focus on the longest flows and does this better
than any other method.  An inversion to recover the original flow distribution
has been attempted in this paper and is partially successful.  Further work in
this area might improve this algorithm.

Packet based sample-and-hold has two advantages.  Firstly, it can be inverted well
to produce a reasonable estimate of the flow distribution even for relatively 
low sampling rates (approximately one in one hundred packets sampled).  Secondly,
it can be resampled to get a packet sample and recover those quantities which
can be measured at the packet, not flow, level.  A disadvantage is that the
estimated probabilities are not guaranteed positive.

SYN based sample-and-hold is near to the original flow-based sampling proposed
by Hohn and Veitch \cite{inverting} but has problems due to packets with more
than one SYN flag.  It is possible that further work could correct for this problem.
However, the problem that this technique will only ever be useful for TCP traffic
remains a major issue.

Our future work on this topic will focus on two main issues.  Firstly, there
is more information to be gained from other parts of the TCP header, notably
the authors know of no results which use the FIN or RST flag for inversion.
While these are problematic since not every flow terminates correctly, still
it would seem that valuable information is contained in these flags.
Secondly, using multiple sample sources could be a rich topic for research.
Some work has already been done in this area:
\cite[Section 6]{fisher} provides a start on this topic focusing on packet
sampling and \cite{gianluca} provides another approach.  
Investigating different sampling techniques which might take advantage
of network topology (for example, if samples are available from two directions
on the same link) could provide more information which might be used to develop
better sampling techniques and also to provide more information for the inversion
problem.

\section*{Acknowledgments}
The authors would like to acknowledge CAIDA \cite{caida} for providing 
the trace files. This work is conducted under the MASTS project 
(EPSRC grant GR/T10503).

%
\bibliographystyle{unsrt}
\bibliography{imc07}  

\begin{thebibliography}{10}

\bibitem{fisher}
Bruno Ribeiro, Don Towsley, Tao Ye, and Jean Bolot.
\newblock Fisher information of sampled packets: an application to flow size
  estimation.
\newblock In {\em IMC '06: Proceedings of the 6th ACM SIGCOMM on Internet
  measurement}, pages 15--26, New York, NY, USA, 2006. ACM Press.

\bibitem{duffield}
Nick Duffield, Carsten Lund, and Mikkel Thorup.
\newblock Estimating flow distributions from sampled flow statistics.
\newblock {\em IEEE/ACM Trans. Netw.}, 13(5):933--946, 2005.

\bibitem{inverting}
Nicolas Hohn and Darryl Veitch.
\newblock Inverting sampled traffic.
\newblock In {\em IMC '03: Proceedings of the 3rd ACM SIGCOMM conference on
  Internet measurement}, pages 222--233, New York, NY, USA, 2003. ACM Press.

\bibitem{choi}
Baek-Young Choi and Supratik Bhattacharyya.
\newblock Observations on cisco sampled net{F}low.
\newblock {\em SIGMETRICS Perform. Eval. Rev.}, 33(3):18--23, 2005.

\bibitem{haddadi}
Hamed Haddadi, Raul Landa, Miguel Rio, and Saleem Bhatti.
\newblock Revisiting the issues on {N}et{F}low sample and export performance,
  2007.
\newblock {\tt arxiv.org/abs/0704.0730}.

\bibitem{estan}
Cristian Estan and George Varghese.
\newblock New directions in traffic measurement and accounting.
\newblock In {\em SIGCOMM '02: Proceedings of the 2002 conference on
  Applications, technologies, architectures, and protocols for computer
  communications}, pages 323--336, New York, NY, USA, 2002. ACM Press.

\bibitem{billing}
Cristian Estan and George Varghese.
\newblock New directions in traffic measurement and accounting: Focusing on the
  elephants, ignoring the mice.
\newblock {\em ACM Trans. Comput. Syst.}, 21(3):270--313, 2003.

\bibitem{netflow}
{Cisco IOS NetFlow}.
\newblock {\tt www.cisco.com}.

\bibitem{Ringberg-Sigmetrics-07}
Haakon Ringberg, Augustin Soule, Jennifer Rexford, and Christophe Diot.
\newblock Sensitivity of {PCA} for traffic anomaly detection.
\newblock Technical Report CR-PRL-2006-11-0001, Thomson lab, 2006.

\bibitem{daniela}
Daniela Brauckhoff, Bernhard Tellenbach, Arno Wagner, Martin May, and Anukool
  Lakhina.
\newblock Impact of packet sampling on anomaly detection metrics.
\newblock In {\em IMC '06: Proceedings of the 6th ACM SIGCOMM on Internet
  measurement}, pages 159--164, New York, NY, USA, 2006. ACM Press.

\bibitem{netflowperf}
Net{F}low performance analysis.
\newblock {\tt www.cisco.com/en/US/tech/tk812/ \\
  technologies\_white\_paper0900aecd802a0eb9 \\ .shtml}.

\bibitem{roughan-comparison}
Matthew Roughan.
\newblock A comparison of {P}oisson and uniform sampling for active
  measurements.
\newblock {\em {IEEE Journal on Selected Areas in Communications}},
  24(12):2299--2312, 2006.

\bibitem{loss}
Robin Sommer and Anja Feldmann.
\newblock Net{F}low: information loss or win?
\newblock In {\em IMW '02: Proceedings of the 2nd ACM SIGCOMM Workshop on
  Internet measurment}, pages 173--174, New York, NY, USA, 2002. ACM Press.

\bibitem{ranking}
Chadi Barakat, Gianluca Iannaccone, and Christophe Diot.
\newblock Ranking flows from sampled traffic.
\newblock In {\em CoNEXT'05: Proceedings of the 2005 ACM conference on Emerging
  network experiment and technology}, pages 188--199, New York, NY, USA, 2005.
  ACM Press.

\bibitem{duffield-learn}
N.~Duffield, C.~Lund, and M.~Thorup.
\newblock Learn more, sample less: Control of volume and variance in network
  measurement.
\newblock Information Theory, IEEE Transactions on, Vol.51, Iss.5, May 2005
  Pages: 1756- 1775.

\bibitem{betternetflow}
Cristian Estan, Ken Keys, David Moore, and George Varghese.
\newblock Building a better {N}et{F}low.
\newblock In {\em SIGCOMM '04: Proceedings of the 2004 conference on
  Applications, technologies, architectures, and protocols for computer
  communications}, pages 245--256, New York, NY, USA, 2004. ACM Press.

\bibitem{willinger1997}
W.~Willinger, M.~S. Taqqu, R.~Sherman, and D.~V. Wilson.
\newblock Self-similarity through high-variability: statistical analysis of
  {Ethernet {LAN}} traffic at the source level.
\newblock {\em {IEEE}\slash{ACM} Trans. on Networking}, 5(1):71--86, 1997.

\bibitem{caida}
{Cooperative Association for Internet Data Analysis (CAIDA)}.
\newblock {\tt www.caida.org}.

\bibitem{flowmon}
Flowmon website.
\newblock {\tt www.flowmon.org/flowmon-probe/flowmon-fw}.

\bibitem{gianluca}
Gion~Reto Cantieni, Gianluca Iannaccone, Chadi Barakat, Christophe Diot, and
  Patrick Thiran.
\newblock Reformulating the monitor placement problem: Optimal network-wide
  sampling.
\newblock 2005.
\newblock Intel Research Technical Report.

\end{thebibliography}
%

\balancecolumns
\end{document}